\documentclass[twocolumn]{aastex7}
\usepackage{hyperref}
\usepackage{amsmath}

\begin{document}

\title{Radio Signature of Higher Atmospheric Meridional Flow and Implications for Magnetic Trees in the Sun}

\correspondingauthor{Anshu Kumari}
\email{anshu@prl.res.in}

\author[0009-0008-5834-4590]{Srinjana Routh}
\affiliation{Aryabhatta Research Institute of Observational Sciences, Nainital-263002, Uttarakhand, India}
\affiliation{Department of Applied Physics, Mahatma Jyotiba Phule Rohilkhand University, Bareilly-243006, Uttar Pradesh, India }
\email{srinjana.routh@gmail.com}

\author[0000-0001-5742-9033]{Anshu Kumari}
\affiliation{Udaipur Solar Observatory, Physical Research Laboratory, Dewali, Badi Road, Udaipur-313001, Rajasthan, India}
\email{anshu@prl.res.in}

\author[0009-0009-1355-5631]{Rohan Bose}
\affiliation{Aryabhatta Research Institute of Observational Sciences, Nainital-263002, Uttarakhand, India}
\affiliation{Department of Physics, Indian Institute of Technology Roorkee, Roorkee-247667, Uttarakhand, India}
\email{rohannewone@gmail.com}

\author[0000-0002-6954-2276]{Vaibhav Pant}
\affiliation{Department of Physics, Indian Institute of Technology, Delhi-110016, India}
\email{vpant@iitd.ac.in}

\author[0009-0001-7689-0084]{Divya Paliwal}
\affiliation{Udaipur Solar Observatory, Physical Research Laboratory, Dewali, Badi Road, Udaipur-313001, Rajasthan, India}
\affiliation{IIT Gandhinagar, Palaj, Gandhinagar, Gujarat, India}
\email{fakeemail@google.com}

\author[0000-0003-4653-6823]{Dipankar Banerjee}
\affiliation{Indian Institute of Space Science and technology, Valiamala, Thiruvananthapuram - 695 547
Kerala, India}
%\affiliation{Indian Institute of Astrophysics, Koramangala, Bangalore 560034, India}
\affiliation{Center of Excellence in Space Sciences India, IISER Kolkata, Mohanpur 741246, West Bengal, India}
\email{dipu@iist.ac.in}

\author[0000-0001-5894-9954]{Nat Gopalswamy}
\affiliation{NASA Goddard Space Flight Center, 8800 Greenbelt Road, Greenbelt, MD 20771, USA}
\email{natchimuthuk.gopalswamy-1@nasa.gov}

\begin{abstract}

The coupling between plasma flows and magnetic fields in the solar atmosphere governs the transport of angular momentum and the redistribution of magnetic flux, yet its manifestation in the magnetically dominated upper chromosphere remains uncertain. Using 27 years of 17 GHz full-disk solar radio imaging observations from the Nobeyama Radioheliograph, we report the first detection of a poleward flow signature at heights of $3000\pm500$ km, an altitude where plasma magnetohydrodynamics expects magnetic dominance ($\beta<1$). The derived latitudinal velocity profile ($5–15$ m/s) mirrors the established photospheric meridional circulation, displaying modulation with solar cycle parameters. Comparison with long-term synoptic magnetograms reveals that the motion of 17 GHz brightness features closely tracks poleward magnetic flux transport, implying a deep magnetic anchoring of these structures. This finding provides the first observational evidence that chromospheric flows at radio wavelengths reflect subsurface meridional dynamics, consistent with the “magnetic tree” hypothesis, which links high-altitude motion to deep-seated magnetic connectivity.
 
\end{abstract}


\section{Introduction} \label{sec:intro}

Large-scale flows in the solar interior and atmosphere regulate the transport of angular momentum and magnetic flux, thereby playing a central role in solar-cycle modulation and magnetic-field evolution. Differential rotation and meridional circulation, in particular, constitute the backbone of flux-transport dynamo models, controlling polar-field reversals, cycle amplitude, and periodicity \citep{Choudhuri1995, Dikpati1999, Zahn1992}. While differential rotation has been extensively characterized across multiple atmospheric layers, with evidence that higher layers can exhibit rotational signatures reflecting deeper anchoring \citep{Badalyan2018, Routh2024}, observational constraints on meridional circulation remain largely confined to the photosphere and convection zone.

Meridional flow, typically observed as a poleward motion of order $10~\mathrm{m\,s^{-1}}$ at the surface \citep{HathawayRightmire2010}, is well established in high-$\beta$ plasma regimes where plasma motions advect embedded magnetic fields. However, the solar atmosphere undergoes a transition from plasma-dominated ($\beta > 1$) conditions in the convection zone and photosphere to magnetically dominated ($\beta < 1$) conditions in the chromosphere \citep{Gary2001}. In low-$\beta$ environments, classical magnetohydrodynamic expectations predict that plasma motions should increasingly align with, rather than govern, magnetic structures. Whether a coherent large-scale meridional flow signature can persist at such heights therefore remains an open question.

From a broader astrophysical perspective, poleward circulations in rotating, stratified, magnetized plasmas arise naturally through baroclinic forcing, Reynolds stresses, and Lorentz-force feedbacks \citep{Miesch2011, Passos2017}. Constraining these flows observationally throughout the solar atmosphere provides critical tests for dynamo theory and informs our understanding of magnetized plasma systems more generally, including other cool stars \citep{petit2018connecting,brun2017magnetism}. Detecting a meridional-flow imprint in the upper chromosphere would therefore establish a direct observational link between deep-seated interior dynamics and magnetically structured atmospheric layers.

Recent work on differential rotation at radio and Extreme Ultraviolet (EUV) wavelengths has suggested that large-scale atmospheric features may remain dynamically connected to deeper layers, consistent with the ``magnetic tree'' hypothesis, in which magnetic structures observed at high altitudes are rooted below the photosphere and reflect the flow profile at their anchoring depths \citep{Weber1969, Routh2024}. If such connectivity extends to meridional circulation, one would expect atmospheric tracers to exhibit poleward migration patterns analogous to those measured at the surface.

In this article, we report the first direct detection of a poleward flow signature at upper-chromospheric heights ($3000 \pm 500$ km) using 27 years of full-disk 17~GHz observations from the Nobeyama Radioheliograph. We show that the derived latitudinal velocity profile closely resembles the established surface meridional circulation in both amplitude and cycle-dependent modulation. Furthermore, comparison with long-term synoptic magnetograms demonstrates that the motion of radio-bright structures tracks poleward magnetic-flux transport. These results provide observational evidence that large-scale meridional dynamics extend into magnetically dominated atmospheric layers, supporting a scenario of deep magnetic anchoring and interior–atmosphere coupling.

\section{Observations and data analysis}\label{sec:data}
In this study, we used Nobeyama Radioheliograph \cite[]{NobeyamaInstrument1994} full-disc solar radio imaging data spanning the declining phase of solar cycle 22, as well as the entire solar cycles 23 and 24. Nobeyama Radioheliograph, operational from 1992 to 2020, has been observing the upper solar chromosphere and lower corona at frequencies of 17 GHz and 34 GHz at a resolution of $10"$. Data beyond 2018 were excluded due to degradation in image quality. 

\begin{figure}
    \centering
    \includegraphics[width=\linewidth]{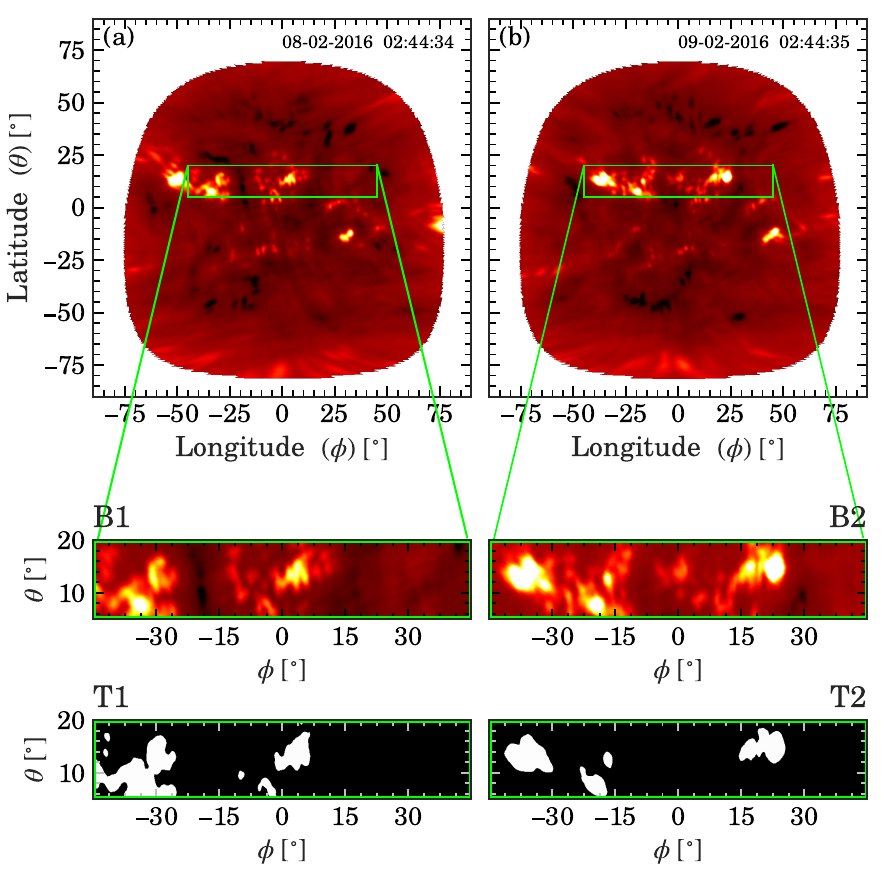}
    \caption{Panels (a) and (b) demonstrate the 17 GHz data from Nobeyama Radioheliograph, temporally separated by 1 day and projected onto heliographic coordinates. Latitudinal bins B1 and B2 are highlighted on which the method of image correlation is applied and thresholded bins T1 and T2 show the dominating features, e.g., active region and bright points,} in the same bin.
    \label{fig:img-corr}
\end{figure}

All images were brightness-normalized and projected into heliographic coordinates \citep{Thompson2006}. The resulting maps were divided into overlapping latitudinal bins of $15^{\circ}$ in width spanning $\pm45^{\circ}$ latitude and $\pm45^{\circ}$ in longitude (\autoref{fig:img-corr}). Such a large bin size, extending from $\theta_1$ till $\theta_2=\theta_1+15^{\circ}$, and their overlapping nature were chosen such that the calculation of cross-correlation coefficient is not biased by the non-uniformity in feature availability \citep{Meunier2003,Riha2007}. For each bin, the image-correlation technique is applied to successive daily maps to determine the shifts along longitude ($\Delta\phi$) and latitude ($\Delta\theta$) for a particular latitude $\theta$, defined as ($\theta=\theta_{mid}=\frac{\theta_1+\theta_2}{2}$), for which the correlation coefficient (C.C.) is maximum.  These shifts are denoted as $u_{\theta}$ for flow along latitude and $\Omega_{\theta}$ for the same along longitude. Since the image-correlation method is independent of tracers and depends only on the intensity pattern of the particular latitudinal bin, the shifts thus calculated are bulk shifts in the same and independent of the biases induced by the distribution of the tracers, as might be in the case of a tracer-dependent method \citep{Olemskoy2005}. This is further discussed in detail for calculating the differential rotation profile using the values of $\Omega_{\theta}$ in \cite{Mishra2024} and \cite{Routh2024}. More recently, this method was tailored to the NoRH data to extract the differential rotational profile in \cite{Routh2025}. 

% In this part, we discuss the values extracted for the latitudinal flow profile. For this study, analysis is limited to $\pm45^{\circ}$, keeping in mind that the bright features contributing to the correlation are majorly limited to this range.

\section{Results}\label{sec:results}

\begin{figure}
    \centering
     \includegraphics[width=\linewidth]{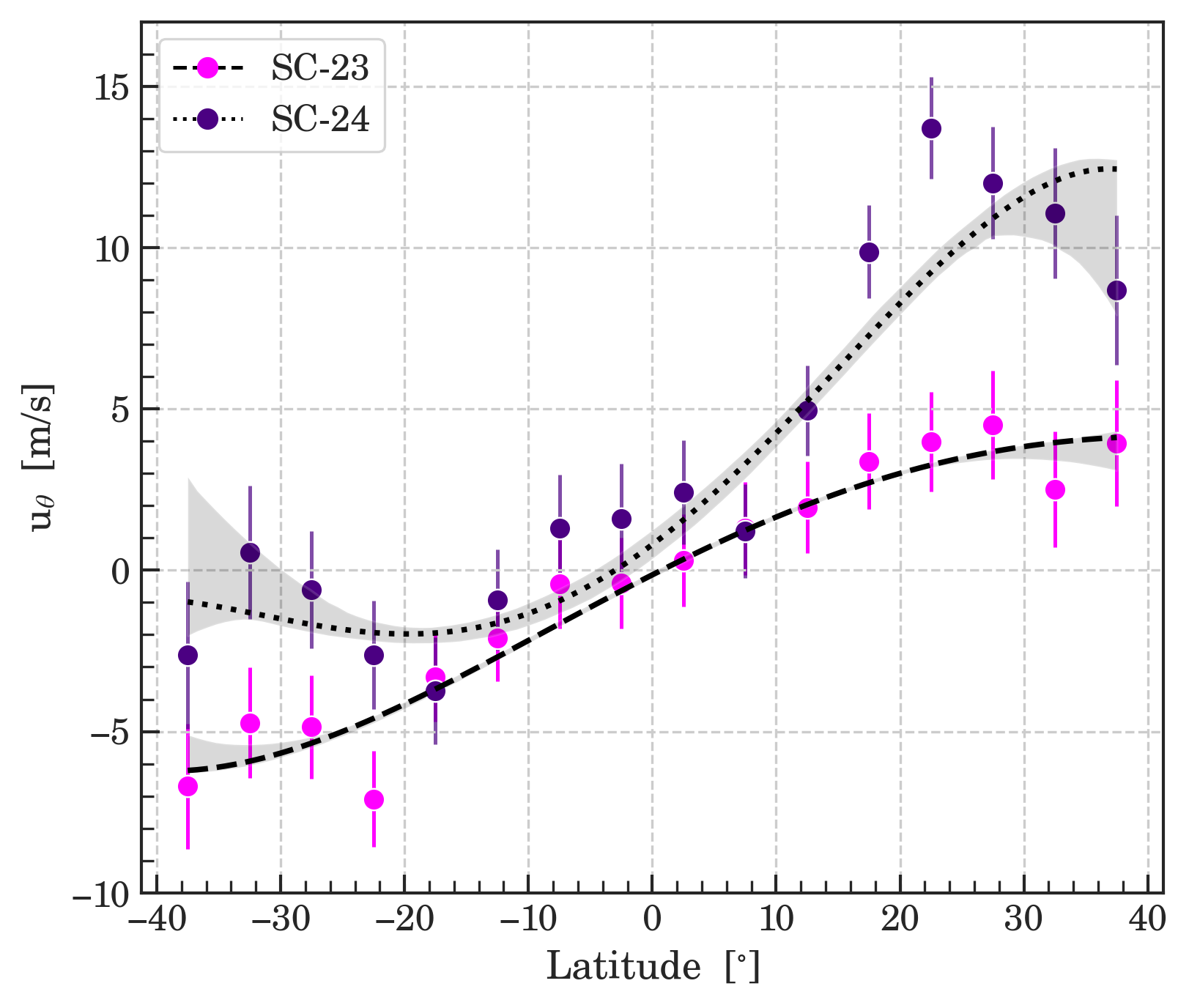}
    \caption{The flow profile signature across solar cycles (SCs) 23 and 24. Error bars signify $3\sigma$ error in the calculation of the point. The black dotted line represents $\chi^2$-optimised fit for that particular dataset, and the dimgray shaded regions signify 95\% confidence interval of the fit.}
    \label{fig:main_flow}
\end{figure}

After setting a threshold on the correlation coefficient (C.C.) based on statistics (see \cite{Mishra2024} for a detailed discussion) to discard values corresponding to very low correlation (in this case, C.C. $<0.35$) for each latitudinal bin, a mean value of the shifts, weighted by their corresponding C.C., was calculated. These shifts were then used to determine the speed $\left(u_{\theta}[^{\circ}/\mathrm{day}]=\frac{\Delta\theta}{\Delta t}\right)$ of the latitudinal flow, which was subsequently converted to linear coordinates to facilitate comparison with photospheric values. Repeating this procedure for all latitudinal bins yielded the latitudinal dependence of the poleward flow, as shown in \autoref{fig:main_flow}.
\subsection{Signature of Poleward Flow}
The average flow profile derived from the combined cycles consistently displays a poleward trend across all analyzed latitudinal bins. Averaged over the declining phase of cycle 22, and the full spans of cycles 23 and 24, the flow maintains an amplitude range of $5–15$ m/s, aligning with established values reported for meridional circulation at the surface and in the convection zone  \citep{HathawayRightmire2010}. A commonly adopted expression, also used in surface flux transport models, expresses the latitudinal variation of the meridional flow in terms of a divergence parameter 
$\Delta_{u}$ and a concentration parameter $p$, in which larger values confine the flow closer to the equator and shift the latitude of maximum speed to $\theta_{peak} = \cos^{-1}\left(\frac{1}{\sqrt{1+p}}\right)$ \citep{Yeats2023}.

 \begin{equation}
    u_{\theta} = -R_{\odot}\Delta_{u}\cos\theta\sin^p\theta
    \label{eq:meridifit}
\end{equation}

Applying this standard, symmetric form to the cycle-averaged profiles derived from the 17 GHz data shows that it does not adequately capture the observed flow, primarily because it enforces north–south symmetry. In particular, it fails to reproduce the differences in peak amplitude and shape between the two hemispheres seen in \autoref{fig:main_flow}. To account for the hemispheric asymmetry, an extended fitting function including an additional parameter $\kappa\sin\theta$ was adopted, where 
$\kappa$ explicitly encodes the degree of asymmetry in the meridional flow profile. 
 \begin{equation}
    u_{\theta} = -R_{\odot}\Delta_{u}\cos\theta\sin^p\theta+\kappa\sin\theta
    \label{eq:meridifit-asym}
\end{equation}
This extra $\kappa$-dependent term allows the latitudinal profile in the north to deviate from that in the south. Helioseismic plasma measurements typically place 
$\Delta_{u}$ in the range $[0.6-1.2]\times10^{-7}$ s$^{-1}$, whereas values inferred directly from meridional flow profiles are often found to be smaller by roughly an order of magnitude (see \cite{Jiang2023}).

The best-fit parameters of this asymmetric flow model were obtained using $\chi^2$minimization, adopting a confidence level of 95\% (see for e.g., \cite{LambStatisticalFit2017}). The optimization was performed with the uncertainties in each bin derived from the scatter of the C.C.-weighted shifts. The resulting best-fit values of all fit parameters for the different cycle and phase ranges covered by the Nobeyama data are summarized in \autoref{tab:phase-based}.

\begin{table*}
\caption{\label{tab:phase-based}The different parameters ($\Delta_u~$, $p$, $\theta_{peak}$, $\kappa$) and their corresponding errors obtained for the best fit of Eq.~\ref{eq:meridifit-asym} to the data.}
\begin{ruledtabular}
\begin{tabular}{cccccc}
Solar Cycle (SC) & Duration/Phase of SC & $\Delta_u~[\mathrm{s}^{-1}]$ & $p$ & $\theta_{peak}$\footnote{The error in $\theta_{peak}$ is obtained on propagating the error for $p$.}& $\kappa$ [m/s] \\  \hline \\[-0.7ex]
  % & & North & South &  & & \\
SC-22 & 1992--1996& $1.9(\pm0.63)\times10^{-8}$ & $0.51(\pm0.3)$ & $\pm35.83(\pm8)^\circ$ & $ -1.74(\pm2.7)$\\[0.7ex]
& Rising & $0.96(\pm 0.41)\times10^{-8}$ & $0.72(\pm0.4)$ &$\pm 41.0(\pm 8.3)^{\circ}$ & $-3.5(\pm 1.4)$\\
 SC-23& Declining & $1.65(\pm 0.44)\times10^{-8}$ & $0.74(\pm0.3)$ & $\pm40.8(\pm5.7)^\circ$ & $4.7(\pm 1.4) $\\
 & 1996--2010 & $1.56(\pm 0.20)\times10^{-8}$ & $0.95(\pm0.18)$ & $\pm44.20(\pm3)^\circ$& $1.76(\pm0.7)$\\[0.7ex]%without bounds
 %  & 1996--2010& $4\pm2$& $7\pm1$ & $0.86\times10^{-8}$ & $0.38$ & $\pm31.65^\circ$\\[0.7ex]
 & Rising & $1.90(\pm 0.38)\times10^{-8}$ & $0.96(\pm0.23)$ &$\pm 44.4(\pm 3.4)^{\circ}$ & $2.8(\pm 0.96)$ \\
 SC-24 & Declining & $1.65(\pm 0.38)\times10^{-8}$ & $0.95(\pm0.3)$ & $\pm 44.3(\pm 6.2)^{\circ}$ & $-6.63(\pm 1) $\\
 & 2010--2018 & $1.87(\pm 0.20)\times10^{-8}$ & $0.82(\pm0.3)$ & $\pm 42.16(\pm 5)^{\circ}$ & $-9.63(\pm 1.6) $\\
\end{tabular}
\end{ruledtabular}
% \footnotesize{*Fit using Eq.~\ref{eq:meridifit} does not converge for these datasets.}
\end{table*}

\subsection{Cycle-wise, phase-wise variations and asymmetry}\label{subsec:cycle_phases}

Inspection of the cycle-separated profiles in \autoref{fig:main_flow} reveals that the overall poleward nature of the flow persists across all analyzed cycles, but the amplitude and degree of asymmetry vary systematically from cycle to cycle. Over cycles 23 and 24, the typical poleward speed remains within the $5–15$m/s range, but the hemispheric peaks and detailed latitude dependence differ. In the two cycles that are fully covered by the present dataset (SC-23 and SC-24), the asymmetry is relatively modest in SC-23 but becomes substantially more pronounced in SC-24. In particular, for SC-24, the cycle-averaged meridional flow peaks at a higher speed in the northern hemisphere than in the southern hemisphere. This difference is reflected quantitatively in the asymmetry parameter $\kappa$, whose absolute value is significantly larger for SC-24 than for SC-23, indicating a stronger north–south imbalance in the flow amplitude.

\begin{figure}[!h]
    \centering
    \includegraphics[width=\linewidth]{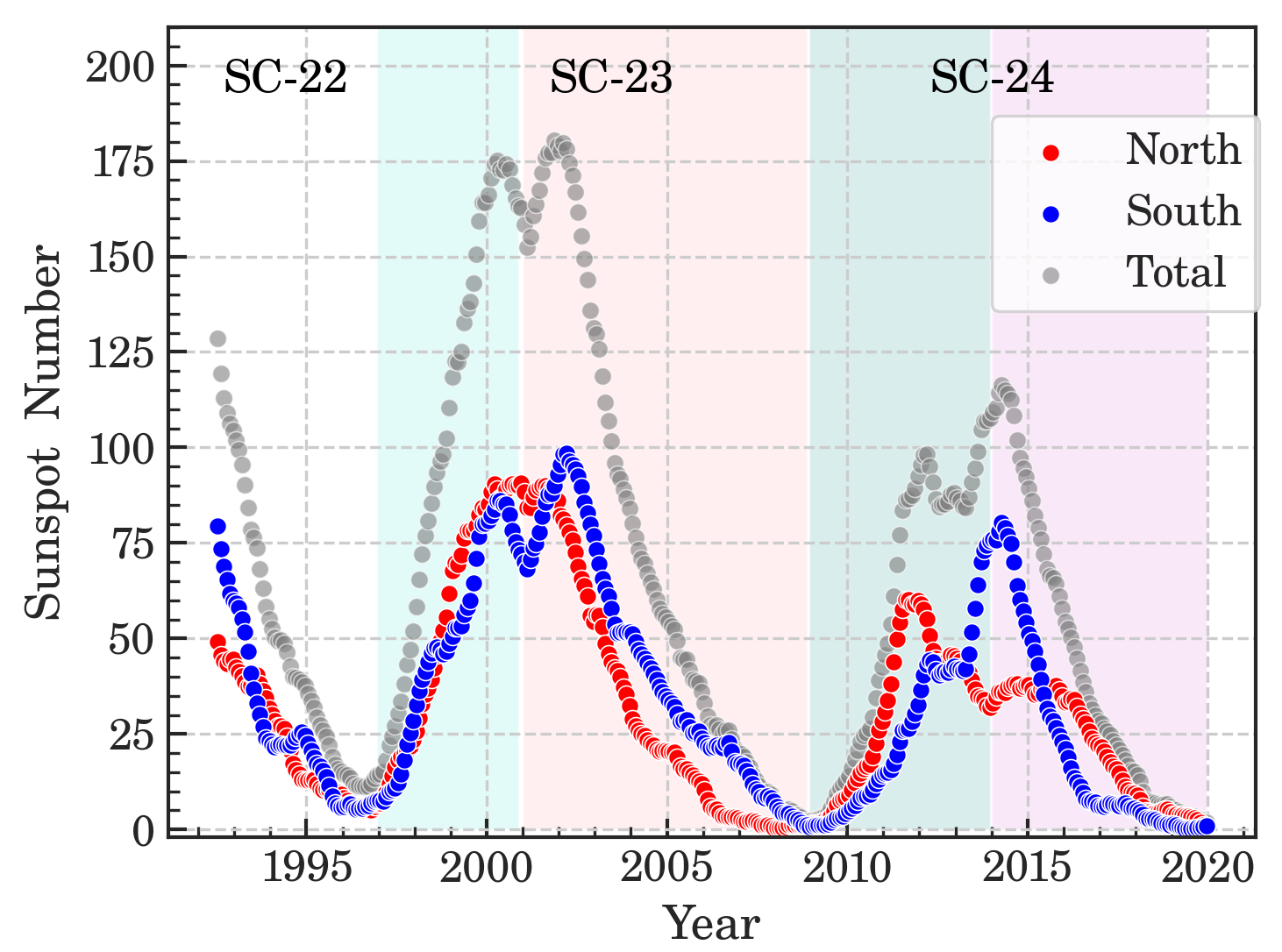}
    \caption{The variation of hemispheric and total sunspot numbers as a marker of solar activity for the entire timespan of observation of NoRH. The rising phases of the cycles 23-24 are highlighted in bluish tones whereas the declining phases in the same cycle are highlighted in pinkish tones, respectively. The SSN data is obtained from the SILSO database \citep{SILSO_Sunspot_Number}. The reader is encouraged to refer \cite{Nat2018Longterm} for a comparison of the same with the polar magnetic field}.
    \label{fig:ssn}
\end{figure}

Such hemispheric asymmetry is in line with earlier results for meridional flows from helioseismology and feature-tracking studies, which have reported hemispheric differences in both amplitude and structure of the subsurface and near-surface meridional flows \citep[see e.g,][]{lekshmi2019hemispheric,Mahajan2021}. A key physical driver of these asymmetries is the localized inflow toward active latitudes, which effectively modulates the background poleward circulation. When one hemisphere exhibits stronger or more persistent magnetic activity, the associated inflows tend to reduce the net poleward speed in that hemisphere. In the period examined here, SC-24 displays systematically stronger activity in the southern hemisphere on average (see \autoref{fig:ssn}), so that the enhanced inflows around active regions there are expected to diminish the cycle-averaged poleward flow speed compared to the north \citep{Imada2020}. This expectation is borne out by the fitted profiles of the flow obtained, in which the northern peak speed remains larger than the southern counterpart, and by the relatively large $|\kappa|$ for SC-24.

\begin{figure*}
    \centering
    \includegraphics[width=0.5\linewidth]{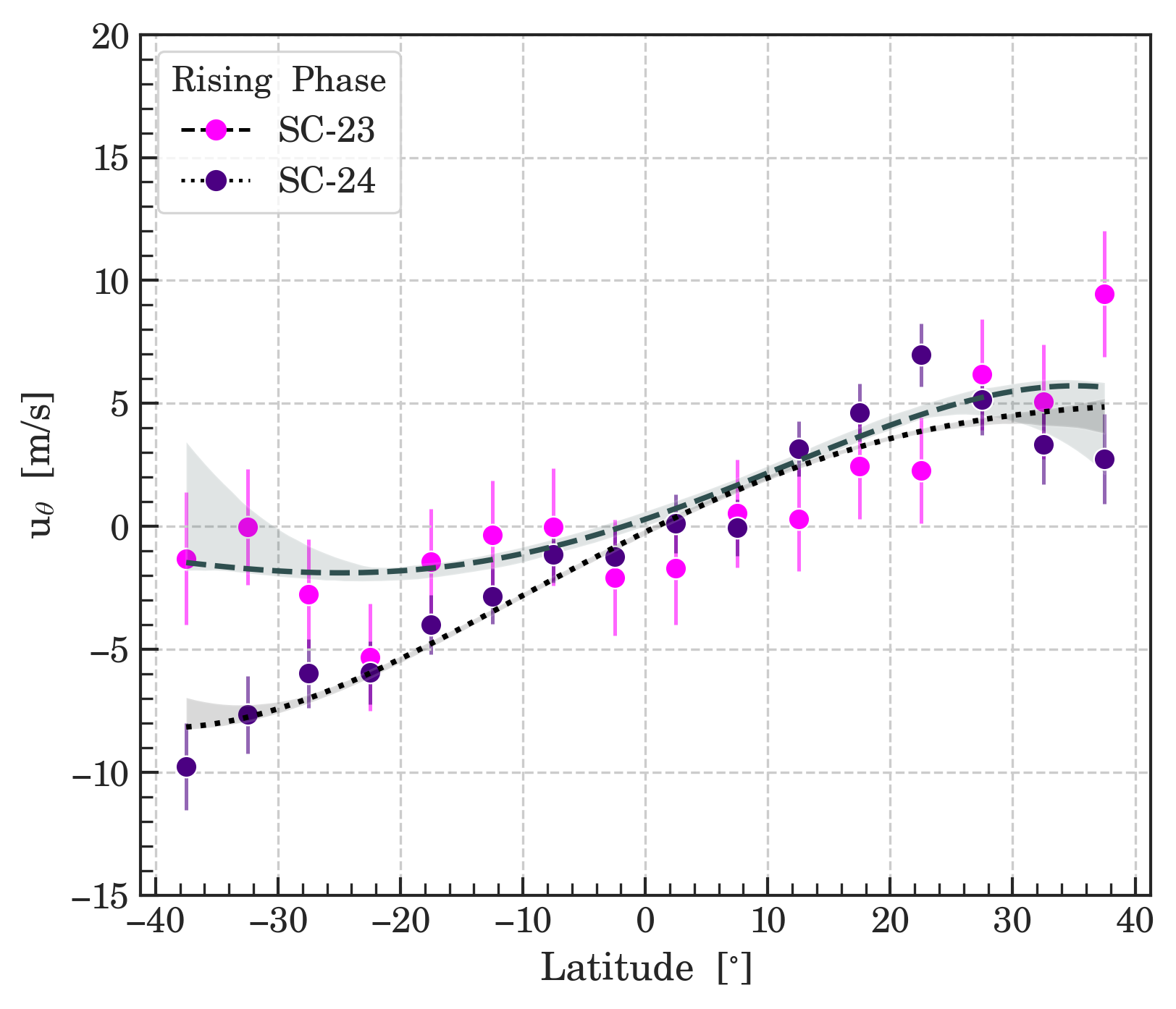}\includegraphics[width=0.5\linewidth]{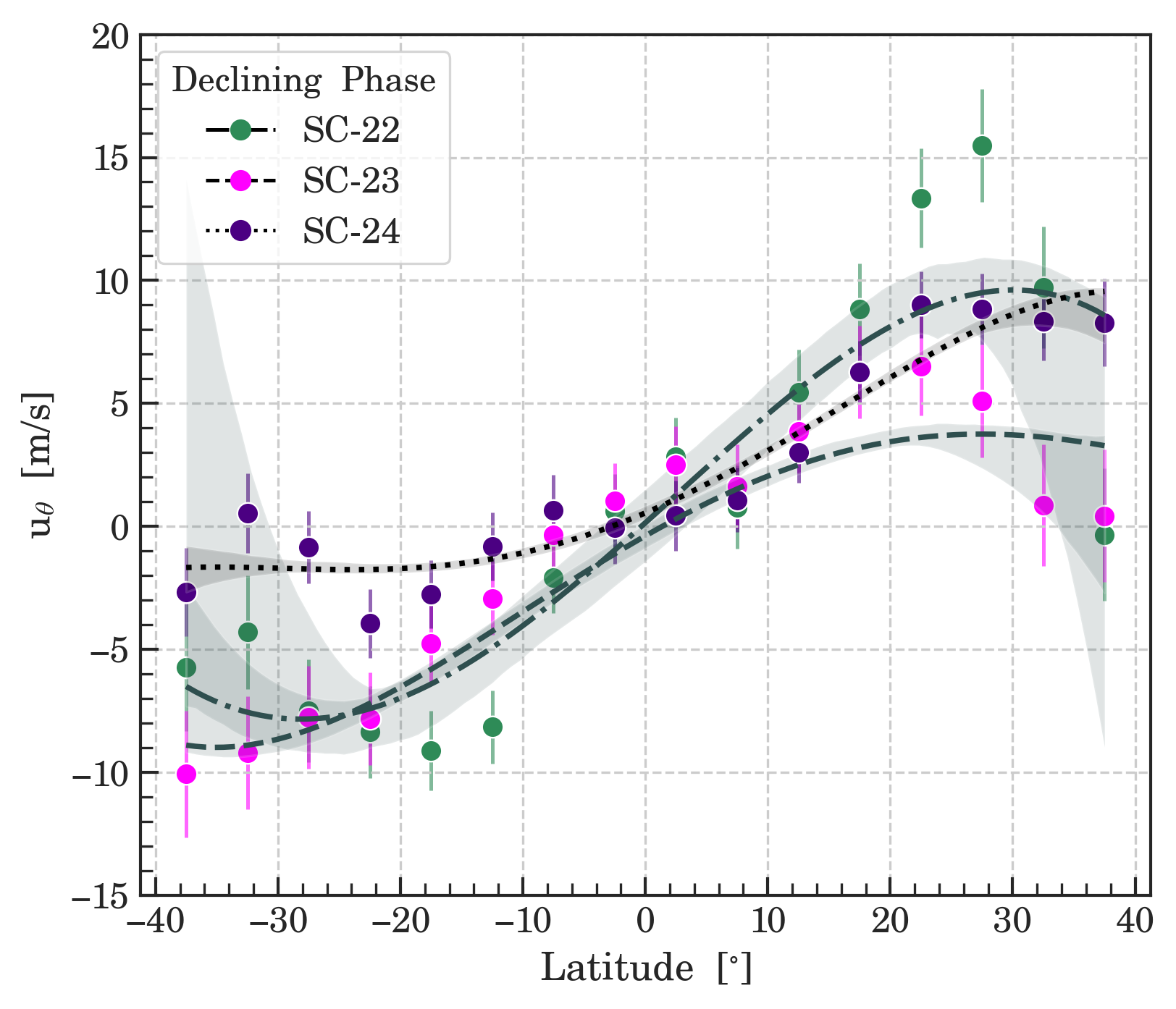}
    \caption{(Left panel) The flow profile in the rising phase of cycles 23 and 24 depicting the poleward flow signature. (Right panel) The same for the declining phase of cycles 22, 23 and 24.}
    \label{fig:phase_based_flow}
\end{figure*}

The asymmetry becomes even clearer when the meridional flow is examined as a function of solar-cycle phase rather than just cycle-averaged. When the data are subdivided into rising and declining phases, defined by the timestamp of the maximum of the corresponding cycle, the phase-separated flow profiles plotted in \autoref{fig:phase_based_flow} show that the departures from north–south symmetry are largest during phases when hemispheric activity is most imbalanced. In those phases where one hemisphere dominates in sunspot number and magnetic flux, the corresponding poleward flow is systematically weaker, consistent with the picture that converging inflows around active regions act to locally slow the background meridional circulation.

In contrast, for SC-23, the cycle-averaged profiles exhibit a smaller degree of hemispheric asymmetry, with the northern and southern peak speeds being more comparable. Nevertheless, there is a clear reduction in the overall amplitude of the cycle-averaged flow in both hemispheres between cycles 23 and 24. This decrease is qualitatively consistent with earlier reports of reduced subsurface meridional speeds in the cycle leading up to the unusually weak polar fields of SC-24 \citep{janardhan2010,Rightmire-Upton2012,Nat2018Polar}. From the perspective of dynamo theory and cycle modulation, such a reduction is also expected: a stronger preceding cycle (SC-23) is associated with enhanced inflows and altered return flows, which in turn can yield a weaker, slower meridional circulation feeding into the subsequent weaker cycle (SC-24). The flow profiles derived here from the 17 GHz data therefore reproduce both the amplitude decline and the hemispheric asymmetry trends that have been suggested as contributors to the weaker polar fields and reduced strength of SC-24.

\subsection{Comparison with magnetogram data}\label{subsec:magnetic_butterfly}

To directly relate the upper-chromospheric flow signatures inferred from the 17 GHz limb-brightening maps to the transport of photospheric magnetic elements, the derived meridional flow profiles were compared with long-term synoptic radial magnetic field maps. The magnetic datasets span the same time interval as the radio observations and include synoptic maps from the Kitt Peak Vacuum Telescope \citep{Livingston1976} (1992–1996), magnetograms from Michelson Doppler Imager \citep{Scherrer1995} (MDI) aboard Solar and Heliospheric Observatory (SOHO) (1996–2010), and Helioseismic Magnetic Imager \citep{Scherrer2012} (HMI) aboard Solar Dynamic Observatory (SDO) (2010–2020). Because these three instruments have different sensitivities and calibration schemes, their mean absolute field strengths in overlapping time intervals were used to derive scale factors, such that MDI fields were scaled relative to Kitt Peak and HMI fields relative to MDI. This inter-calibration ensures that the combined synoptic map set forms a homogeneous time series suitable for quantitative comparison with the radio data.

% We overlaid intensity (brightness temperature, $T_B$) contours extracted from the 17 GHz limb-brightening maps, which were generated following a similar method described in \cite{Selhorst2003}. Contour levels of $12,000$ K and $11,300$ K delineate regions of enhanced activity near low latitudes and polar brightening at high latitudes or brightenings in polar coronal holes , respectively. Remarkably, the temporal and spatial evolution of the radio features shows a close correspondence with the evolution of magnetic structures observed in magnetograms, particularly in latitude-time domains associated with meridional flow. Notably, the congruent overlap in SC-23 from approximately 2002 and in SC-24 from 2014 underscores that the motion of features identified in the 17 GHz data faithfully tracks the poleward migration of magnetic flux, which is a key signature of meridional circulation. This suggests that the motion of features captured through the analysis of 17 GHz data indeed coincides with the motion of magnetic fields, as has been previously explored in \citep{Nat2018Polar}.

\begin{figure*}
    \centering
    \includegraphics[width=\linewidth]{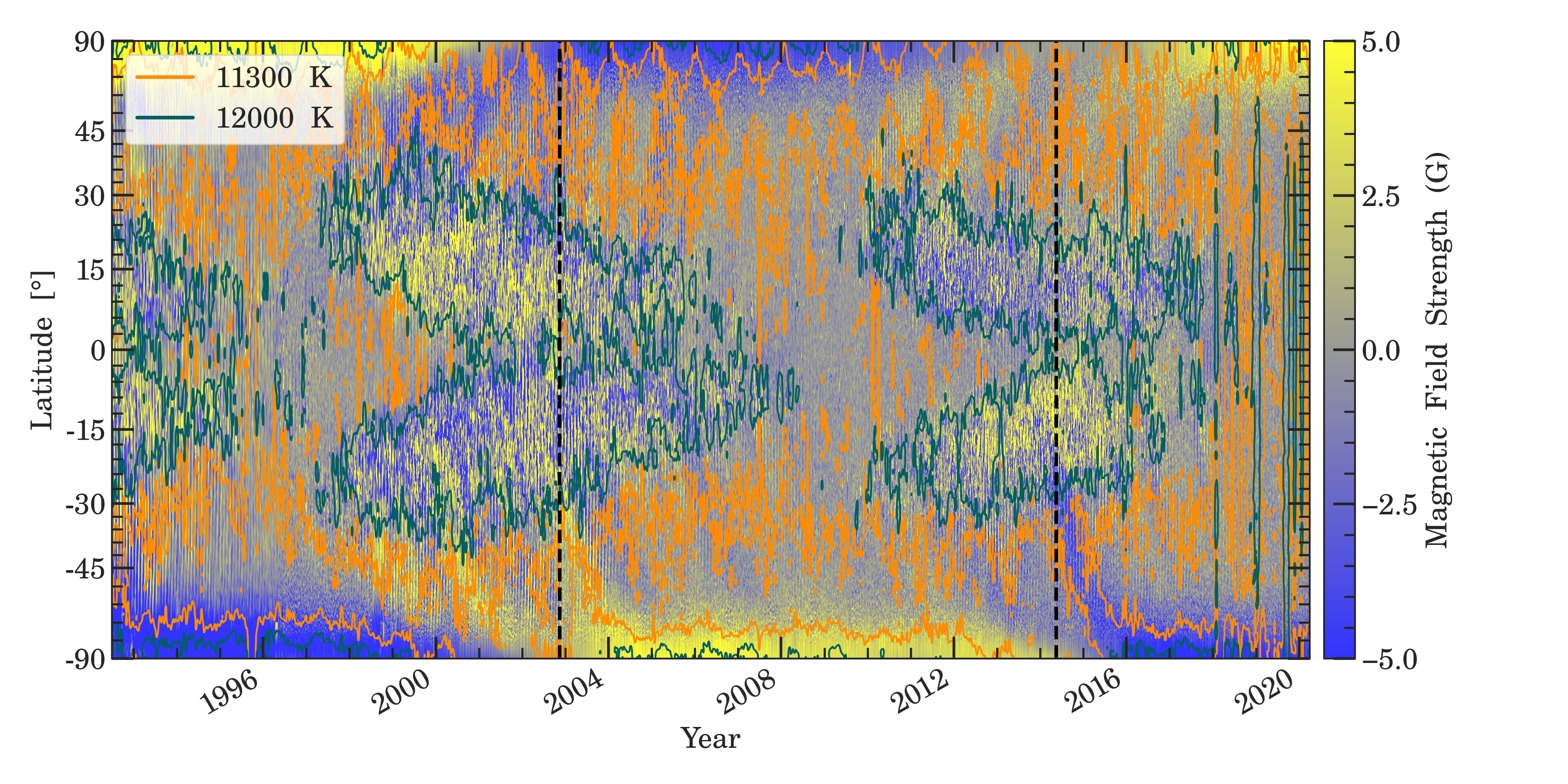}
    \caption{$T_B$ contours overplotted on the magnetic butterfly diagram. The two contour levels primarily correspond to active regions at low latitudes and polar brightenings at high latitudes. The black vertical dashed lines indicate the locations where the meridional flow signature first appeared in both the radio and magnetic field maps, clearly distinguishable in the southern hemisphere.}
    \label{fig:magnetic butterfly}
\end{figure*}

From the Nobeyama 17 GHz observations, limb-brightening maps were processed following an approach similar to that of \cite{Selhorst2003} to extract equal brightness temperature ($T_B$) contours corresponding to structures of enhanced activity. Two particular contour levels were selected for detailed comparison: $T_B$ of $12,000$ K, which traces regions of enhanced activity at low to mid latitudes, and $T_B$ of $11,300$ K, which highlights polar brightenings \citep[see e.g.,][]{Nat1998}. These radio-brightness contours were overlaid on the latitude–time diagrams of the inter-calibrated radial magnetic field maps, thereby enabling a direct visual and quantitative comparison between the temporal evolution of the radio features and that of the magnetic flux (see \autoref{fig:magnetic butterfly}).

The overlaid plots reveal a striking correspondence between the radio and magnetic structures, similar to \cite{Nat2018Longterm}. Additionally, in the low-latitude belts associated with active regions and in the high-latitude zones where poleward-migrating flux accumulates, the 17 GHz brightenings closely follow the same latitude–time tracks as the magnetic features. This agreement is particularly evident during solar cycle 23 from about 2002 onward and during solar cycle 24 from around 2014, where the 17 GHz bright contours and the bands of enhanced radial field remain nearly co-spatial as they drift toward the poles. Incidentally, this also coincides with the rush to the poles (RTTP) of the polar filaments as discussed in \cite{Nat2012}. Such a congruent evolution in the latitude–time domain strongly suggests that the features tracked in the 17 GHz data are tied to the same large-scale meridional circulation that transports magnetic flux poleward at the photosphere.

\section{Discussion and conclusion}\label{sec:conc}

In this study, we analyze the large-scale flow along the latitudinal plane at a height of $3000\pm500$ km in the upper chromosphere \citep{zirin1988astrophysics} using 27 years of 17 GHz Nobeyama Radioheliograph data, covering entirety of cycles 23 and 24 and part of cycle 22. This analysis provides, to our knowledge, the first clear evidence of a poleward flow at upper chromospheric heights whose large-scale behavior closely resembles well-established subsurface and near-surface meridional flows. The flow amplitude decreases with increasing cycle strength and exhibits a pronounced hemispheric asymmetry that is anti-correlated with the hemispheric asymmetry in activity, consistent with known properties of meridional circulation.

To assess whether the observed flow trend in the upper chromosphere is consistent with meridional flow signatures inferred from photospheric magnetograms, we conducted a comparative analysis of the temporal evolution of features that exceed the quiet Sun brightness temperature (\textbf{$T_B$} = 10,000 K). Brightness enhancements in this range, predominantly bright points and small-scale active regions, were identified in the 17 GHz maps and cross-compared with magnetic structures detected in magnetograms from multiple instruments. The two datasets display a striking correspondence in their latitude–time evolution, particularly in the domains where meridional transport is expected to dominate, indicating that the chromospheric flow measured at 17 GHz reliably traces the underlying transport of photospheric magnetic flux by meridional circulation. It is worth noting here that the high-latitude open magnetic field correlates very well with the $T_B$ of NoRH, thereby adding to the significance of the flow in the context of the global magnetic field structure of the Sun \citep{Fujiki2019}.

The presence of such a flow signature at a height where the solar atmosphere is magnetically dominated ($\beta<1$) stands in contrast to classical expectations, where plasma flow dictates magnetic field transport only when $\beta>1$. This suggests that the flow may not be governed solely by local plasma conditions at the observed height. Instead, these magnetic structures may remain dynamically anchored to deeper layers,  in line with the magnetic tree scenario and previous proposals that attribute angular-momentum transport and rotational shear to deep-rooted magnetic connectivity \citep{Weber1969}. Previous observations of large-scale features at 17 GHz radio wavelengths, as well as in 304 {\AA} and 171 {\AA} Extreme Ultraviolet (EUV) bands, have similarly revealed rotational profiles mirroring those of the solar interior \citep{Routh2024,Routh2025}. This might suggest that the upper atmosphere reflects not only differential rotation but also meridional circulation of the interior plasma through the large-scale atmospheric features anchored in it. Such a configuration may explain both the high-atmosphere reflection of deep meridional flows and the occurrence of faster atmospheric rotation relative to the photosphere. A detailed understanding of the origin of this flow, however, remains outside the scope of this study.

While these results open a new observational window on the coupling and dynamics of the solar atmosphere using radio data, the present analysis is limited to two complete activity cycles and thus does not yet permit a comprehensive parametric study of cycle-to-cycle variability. Also, due to the availability restrictions of the features analyzed in this study, the analysis is limited to active region belts only. Extending this approach to longer time series as well as higher latitudes in future work will be essential for constraining the properties of deep magnetic structures and their anchoring throughout the atmosphere, and for incorporating these constraints into more predictive models of the solar and stellar dynamo as well as large-scale flow evolution.

\begin{acknowledgments}
The data utilized in this work were acquired from the Nobeyama Radioheliograph (NoRH) database. The authors are grateful to the International Consortium for the Continued Operation of Nobeyama Radioheliograph (\href{https://hinode.isee.nagoya-u.ac.jp/ICCON/}{ICCON}) for making the NoRH data available on line. This work has benefited from NASA's open data policy for SDO/HMI and SOHO/MDI data as well as from the SILSO database. NSO/Kitt Peak data used in this study are produced cooperatively by NSF/NOAO, NASA/GSFC, and NOAA/SEL. The ISEE Database for High-Cadence Microwave Images of Solar Flares Observed with Nobeyama Radioheliograph was developed by the Center for Heliospheric Science, Institute for Space-Earth Environmental Research (ISEE), Nagoya University. S.R. is supported by funding from the Department of Science and Technology (DST), Government of India, through the Aryabhatta Research Institute of Observational Sciences (ARIES). The computational resources utilized in this study were provided by ARIES. AK acknowledges the ANRF Prime Minister Early Career Research Grant (PM ECRG) program. RB acknowledges the financial support received from the DST-INSPIRE Fellowship Programme (DST/INSPIRE Fellowship/2021/IF210402). NG is supported by NASA's LWS program and the STEREO project. A brief note of gratitude is extended to Bibhuti Kumar Jha for insightful comments at the initial stages of analysis.
\end{acknowledgments}

\begin{contribution}
All authors have made substantial contributions to the work reported in this manuscript.

\end{contribution}
\software{IDL, scipy, numpy, OpenCV, sunpy, pandas, seaborn.}
\bibliography{references}{}

\bibliographystyle{aasjournal}

\end{document}